\documentclass[12pt]{article}
\usepackage{pdproc,epsfig} 

\begin{document}

\def\beq{\begin{equation}}
\def\eeq{\end{equation}}
\def\bea{\begin{eqnarray}}
\def\beaa{\begin{eqnarray*}}
\def\eea{\end{eqnarray}}
\def\eeaa{\end{eqnarray*}}
\def\bq{\begin{quote}}
\def\eq{\end{quote}}
\def\gappeq{\mathrel{\rlap {\raise.5ex\hbox{$>$}}
{\lower.5ex\hbox{$\sim$}}}}

\def\lappeq{\mathrel{\rlap{\raise.5ex\hbox{$<$}}
{\lower.5ex\hbox{$\sim$}}}}
\def\bigP{\mbox{\boldmath$P$}}
\def\bigR{\mbox{\boldmath$R$}}
\parskip 0.3cm
    \newcommand{\dotg}{\!\!\!/}     \newcommand{\pslash}{p_{\!}\!\!/}
    \newcommand{\dd}{{\rm d}}       \newcommand{\td}{\!{\rm d}}
     \newcommand{\nslash}{n_{\!}\!\!/}                               

\renewcommand{\thefootnote}{\alph{footnote}}
  
\title{A NEUTRINO FACTORY MANIFESTO}

\author{John ELLIS}

\address{ Theoretical Physics Division, CERN\\
 CH-1211 Gen\'eve 23, Switzerland\\
 {\rm E-mail: John.Ellis@cern.ch}}

  \centerline{\footnotesize CERN-TH/2001-142~~~~~~hep-ph/0105265}

\abstract{A neutrino factory is capable of answering basic
questions in the physics of neutrino oscillations, which address
fundamental issues in Grand Unified Theories and flavour physics. In
addition, the front end of a neutrino factory offers exciting prospects
using slow and stopped muons, such as the search for $\mu \rightarrow e$
transitions and the muon electric dipole moment. There are also
opportunities to study muonic atoms and in other areas of science. One
should also keep in mind the long-term objective of building a muon
collider, either for Higgs physics or at the high-energy frontier.}
   
\section{Neutrino Masses and Oscillations}

Neutrino masses and oscillations offer unique perspectives on physics
beyond the Standard Model. There was never any good reason why neutrino
masses should vanish, there being no exact gauge symmetry to prevent them
from acquiring masses. In the cases of the photon and gluons, the exact
$U(1)$ and $SU(3)$ gauge symmetries of the Standard Model safeguard their
masslessness. There is no corresponding massless gauge boson coupling to
lepton number, so theorists have long being expecting non-zero neutrino
masses~\cite{numasses}. 

There are many models of neutrino masses based on GUTs~\cite{AF},
theories of flavour with additional $U(1)$ generation
symmetries~\cite{U1}, and recently models with extra
dimensions~\cite{extraD}. In general, one may say that neutrino masses
open a window directly on physics at a high mass scale, possibly beyond
the reach of collider experiments, providing us with a look at GUTs,
flavour physics, and perhaps even quantum gravity~\cite{QG}.

The simplest form of neutrino mass term is the Majorana type
$m_\nu \nu \nu$, which could even be generated within the Standard Model,
via non-renormalizable interaction of the form~\cite{BEG}
\begin{equation} 
{1 \over M} \nu H \cdot \nu H \rightarrow m_\nu = { \langle 0 | H |0
\rangle^2 \over M}
\label{BEG}
\end{equation} 
where $M$ is some heavy mass scale $ \gg m_W$, and $H$ denotes the
Standard Model Higgs field. This mechanism naturally gives $m_\nu \ll
m_{q, \ell}$, as is suggested by experiment, but begs the questions where
$M$ originates. 

The favoured origin of $M$ is in some Grand Unified Theory (GUT) see-saw
model~\cite{AF}, which looks like
\begin{equation} 
(\nu_L , \nu_s )
\left (
\begin{array}{cc} 
0 & m_0\\  
m_0^T & M_M   
\end{array}
\right )
\left (
\begin{array}{c}
\nu_L\\
\nu_s
\end{array}
\right )
\label{seesaw}
\end{equation} 
in its simplest form, where one postulates a singlet neutrino $\nu_s$ with
a large Majorana mass $M_M$, and the Dirac mass $m_D = {\cal O}(m_{q,
\ell})$.  After diagonalization, the matrix (\ref{seesaw}) yields a light
mass eigenvalue:
\begin{equation} 
m_\nu = m_D {1 \over M_M} m_D^T.
\label{lightmass}
\end{equation}
Both the equations (\ref{lightmass}) and (\ref{seesaw}) should actually be
regarded as matrices in flavour space. After their diagonalization, and
that of the charged-lepton mass matrix, there will in general be a
mismatch in flavour space, that is interpreted as the neutrino mixing
matrix: 
\begin{equation} 
V_{MNS} \equiv V_\ell V_\nu^\dagger
\label{MNS}
\end{equation} 
In view of the very different origin (\ref{lightmass}) from that of quark
masses, involving the heavy Majorana mass matrix $M_M$ as well as the
Dirac matrix $m_D$, it should not be surprising if neutrino mixing is very
different from that of quarks: $V_{CKM} = V_d V_u^\dagger$. 

The see-saw framework (\ref{seesaw}) may well be a gross simplification of
the new physics of neutrino masses. For example, in some models with extra
dimensions~\cite{extraD}, there are an infinite sequence of excited
Kaluza-Klein states coupled to the familiar $\nu_L$. In consequence, there
may be observable deviations from naive sinusoidal oscillation patterns
{\it in vacuo}~\cite{nonsine}, and there may be multiple MSW effects in
matter~\cite{multiMSW}.  Intermediate between these Kaluza-Klein models
and the minimal see-saw (\ref{seesaw}) are some string-inspired (or
­derived) models in which there are a large but finite number of massive
states mixing with the $\nu_L$. More exotic possibilities suggested by
some quantum theories of gravity include violations of the equivalence
principle~\cite{EP} or Lorentz invariance~\cite{LI}, and quantum
decoherence~\cite{QG}. 

\section{The Emerging Default Option}

As discussed by many speakers at this meeting, there is much confirmed 
evidence for both atmospheric neutrino oscillations, with~\cite{Atmo}
\begin{equation}
10^{-2}~{\rm eV}^2 \gappeq \Delta m^2_{Atmo} \gappeq 10^{-3}~{\rm eV}^2,
\label{Atmo}
\end{equation}
and solar neutrino oscillations with~\cite{Solar}
\begin{equation}
10^{-4}~{\rm eV}^2 \gappeq \Delta m^2_{Solar} \gappeq 10^{-11}~{\rm eV}^2,
\label{Solar}
\end{equation}
There is also evidence from the LSND experiment for oscillations with a
larger value of $\Delta m^2$~\cite{LSND}, that cannot be accommodated
within any
simple three-generation scenario, but could be explained with an
additional light sterile neutrino $\nu_s$. If the LSND result were to be
confirmed~\cite{MiniB}, neutrino
oscillation physics would be even more interesting. However, in the rest
of this review, {\it we shall stick our heads in the sand and be very
conservative, restricting ourselves to the seesaw model (\ref{seesaw}) and
just three light neutrino species}, with $\Delta m^2$ in the ranges
(\ref{Atmo}, \ref{Solar}). The geometry of three-flavour mixing may be
described by $(\nu_e, \nu_\mu, \nu_\tau)^T = U \cdot (\nu_1, \nu_2,
\nu_3)^T$,
where
\begin{equation} 
U = \left (
\begin{array}{ccc}
1 & 0 & 0 \\
0 & c_{23} & s_{23} \\
0 & -s_{23} & c_{23}
\end{array} \right )
\left ( \begin{array}{ccc}
c_{13} & 0 & s_{13}e^{i \delta} \\
0 & 1 & 0 \\
-s_{13}e^{-i \delta} & 0 & c_{13} \end{array} \right )
\left ( \begin{array}{ccc}
c_{12} & s_{12} & 0 \\
-s_{12} & c_{12} & 0 \\
0 & 0 & 1
\end{array} \right)
\label{Euler}
\end{equation} 
with three (Euler) mixing angles $\theta_{23,13,12}$ and one
CP-violating phase $\delta$. 

There are upper limits on neutrino masses from astrophysics and cosmology,
which suggest that all three light neutrino flavours weigh less than about
2~eV~\cite{cosmo}, and Tritium $\beta$ decay~\cite{beta}, which suggests
independently that the
state which is predominantly $\nu_e$ weighs less than about 2~eV. Hence,
we have
\begin{equation} 
m_\nu \lappeq 2~{\rm eV} \gg \sqrt{\Delta m^2_{Atmo}} \gg 
\sqrt{\Delta m^2_{Solar}}.
\label{hierarchy}
\end{equation} 
One may ask whether all three neutrino species might have (almost) 
degenerate masses close to the upper limit $\sim 2$~eV. If so, there must
be strong cancellations to ensure that $\langle m_{\nu_e} \rangle \lappeq
0.2$~eV, as required by neutrinoless double-$\beta$
decay~\cite{doublebeta}. This requires
almost bimaximal mixing, which is possible if solar neutrino data are
described by vacuum oscillations (VO), but not if the small-mixing-angle
(SMA) solution is correct, and perhaps not even in the large-mixing-angle
(LMA) case, since this is valid only if $\sin^2 2 \theta_{12}$ is bounded
away from unity. However, in the surviving VO case, one would require
$m_\nu \simeq 10^{10} \times \sqrt{\Delta m^2_{Solar}}$, which would be
quite impressive, and totally unexpected within a conventional see-saw
model. Moreover, such degeneracy and bimaximal mixing are difficult to
reconcile with the inevitable renormalization of neutrino masses at
sub-GUT scales~\cite{EL}. {\it Therefore, we disfavour degenerate neutrino
scenarios
in the following.}

As is well known, the most likely scenario for interpreting the
atmospheric neutrino data is that $\nu_\mu \to \nu_e$ oscillations are not
dominant, because of upper limits from the Super-Kamiokande~\cite{Atmo}
and Chooz experiments~\cite{Chooz}.  Nor can $\nu_\mu \to \nu_{s}$
oscillations be dominant, because of the azimuthal-angle distributions
observed by Super-Kamiokande~\cite{Atmo}. Thus, near-maximal $\nu_\mu
\to \nu_\tau$ oscillations are favoured. Moreover, Super-Kamiokande has
recently reported~\cite{SKnutau} the possible detection of $\tau$
production at the 2-$\sigma$ level, close to the expected rate. There are
more prospects for confirmation of $\nu_\mu \to \nu_\tau$ oscillations by
MINOS~\cite{MINOS}, and final proof is likely to come from
OPERA~\cite{OPERA}. {\it We assume as a default that $\theta_{23}$ is
large, whilst $\theta_{13}$ is small.}

In the case of solar neutrinos, the rates alone do not distinguish between
the LMA, SMA, VO and low-mass (LOW) solutions. However, the day-night
energy spectra favour LMA~\cite{Solar}, though not yet conculsively. 
Since the physics case for the neutrino factory is strongest in the LMA
case, we should avoid jumping to favourable conclusions!  Definitive
answers may soon be provided by the SNO~\cite{SNO} and KamLAND~\cite{KL}
experiments, and by Borexino~\cite{BO}. However, for the time being, {\it
we assume the LMA solution, with $\theta_{12}$ large and $\Delta m^2_{12}
\sim {\rm few} \times 10^{-5}$~eV$^2$ to $10^{-4}$~eV$^2$}, even though
this seems almost too good to be true! 

Thus, the emerging default option for neutrinos comprises three light
neutrinos with hierarchical masses and (almost) bimaximal mixing: 
\beq
V_{\nu} \simeq \left(
\begin{array}{ccc}
\frac{1}{\sqrt{2}} & -\frac{1}{\sqrt{2}} & 0 \\
\frac{1}{2} & \frac{1}{2} & -\frac{1}{\sqrt{2}} \\
\frac{1}{2} & \frac{1}{2} & \frac{1}{\sqrt{2}}
\end{array}
\right)
\label{7}
\eeq
Their masses are thought to be effectively mainly of Majorana nature, they
are expected to have small dipole moments, and their lifetimes are
expected to be much greater than the age of the Universe. 

This brief summary begs many big issues.  Can we exclude the existence of
one or more light sterile neutrinos $\nu_s$? Can we exclude (almost)
degenerate neutrino masses or an inverse mass hierarchy?  If mixing is
indeed (nearly) bimaximal (7), how can we discriminate between the LMA and
LOW solutions, and how big is $\theta_{13}$?  Are we really so lucky that
CP violation is observable in neutrino oscillations? If the neutrino
masses really are Majorana, can we fix them by neutrinoless double-$\beta$
decay experiments?  If the anomalies in the solar and atmospheric data are
indeed due to oscillations, rather than decays, can we see the oscillation
pattern? 

\section{Programme of Work for a $\nu$ Factory}

There is a very full programme of work for a neutrino factory, with its
centrepiece being neutrino oscillation studies~\cite{PH}.  Among the
experimental objectives are determining the magnitude of $\theta_{13}$,
observing MSW matter effects under controlled conditions~\cite{matter} and
determining the sign of $\Delta m^2_{23}$, and measuring the CP violation
phase $\delta$.  Theoretical details of this programme are discussed here
by Pilar Hernandez~\cite{PH}, so here I just note a few points. 

\begin{figure}[htb]

\begin{center}
\hspace{-0.5in}
\epsfig{figure=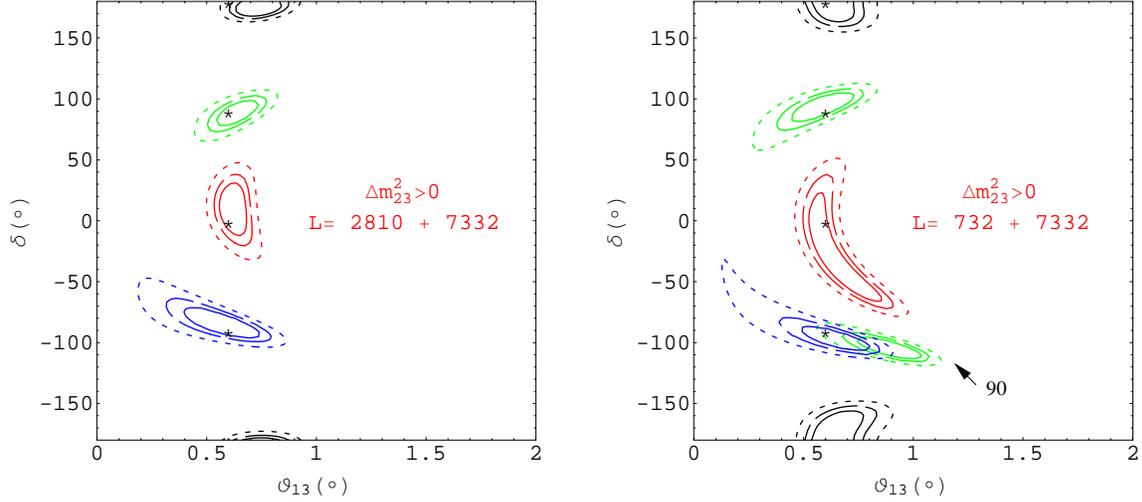,height=3in} 
\end{center}
\caption{\it Estimates of the precisions possible in measurements of
$\theta_{13}$ and the CP-violating phase $\delta$ at a
$\nu$ factory~\protect\cite{combine}, combining data from two different
long
baselines.}
\label{fig:combine}
\end{figure}

\begin{figure}[htb]

\begin{center}
\epsfig{figure=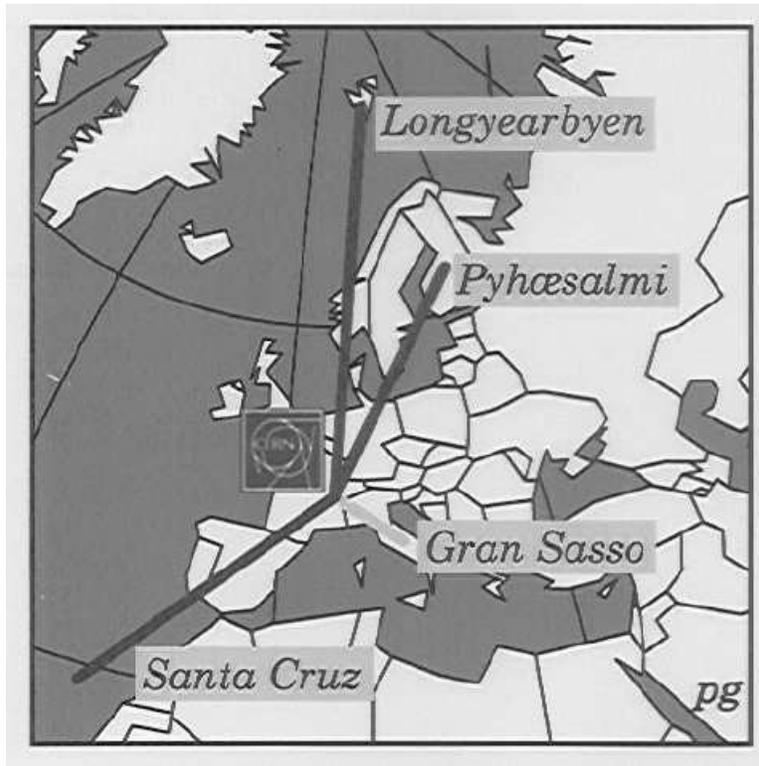,height=4in} 
\end{center}
\caption{{\it Possible locations of long-baseline neutrino detectors that
could make measurements using beams from a $\nu$ factory at
CERN~\protect\cite{HW}.}}
\label{fig:Horst}
\end{figure}

The optimal single distance for CP-violation studies is about 3000 to 4000
km~\cite{best3000}.  However, the precision may be improved by combining
experiments at different baselines~\cite{combine},
as seen in Fig.~\ref{fig:combine}. For this reason,
CERN and other laboratories have been looking for sites suitable for
experiments at distance around 3000 km. There are not so many suitable
places in Europe that are so distant from CERN, but they do include the
Santa Cruz de Tenerife in the Canary Islands - where there are many
tunnels in the volcanic rock - Lonyearbyen in Spitzbergen - where there
are coal mines - and Pyh\"asalmi in Finland - where there is a mine for
zinc and copper, as seen in Fig.~\ref{fig:Horst}~\cite{HW}.

In designing a neutrino factory, other physics opportunities should be
kept in mind.  These include an intense low-energy $\nu$ 
`superbeam'~\cite{superbeam},
physics with slow (or stopped) 
muons~\cite{GG} - such as flavour-violating processes
including $\mu \to e \gamma$ decay and $\mu \to e$ conversion on a heavy
nucleus, a new measurement of $(g_{\mu} -2)$,
deep-inelastic $\nu$ (or $\mu$?)  scattering~\cite{MM}, neutron physics,
physics with radioactive beams, kaon physics, etc.  Some of these are
described in more detail later, but first let us acquaint ourselves better
with the conceptual design of a $\nu$ factory. 

\begin{figure}[htb]

\begin{center}
\epsfig{figure=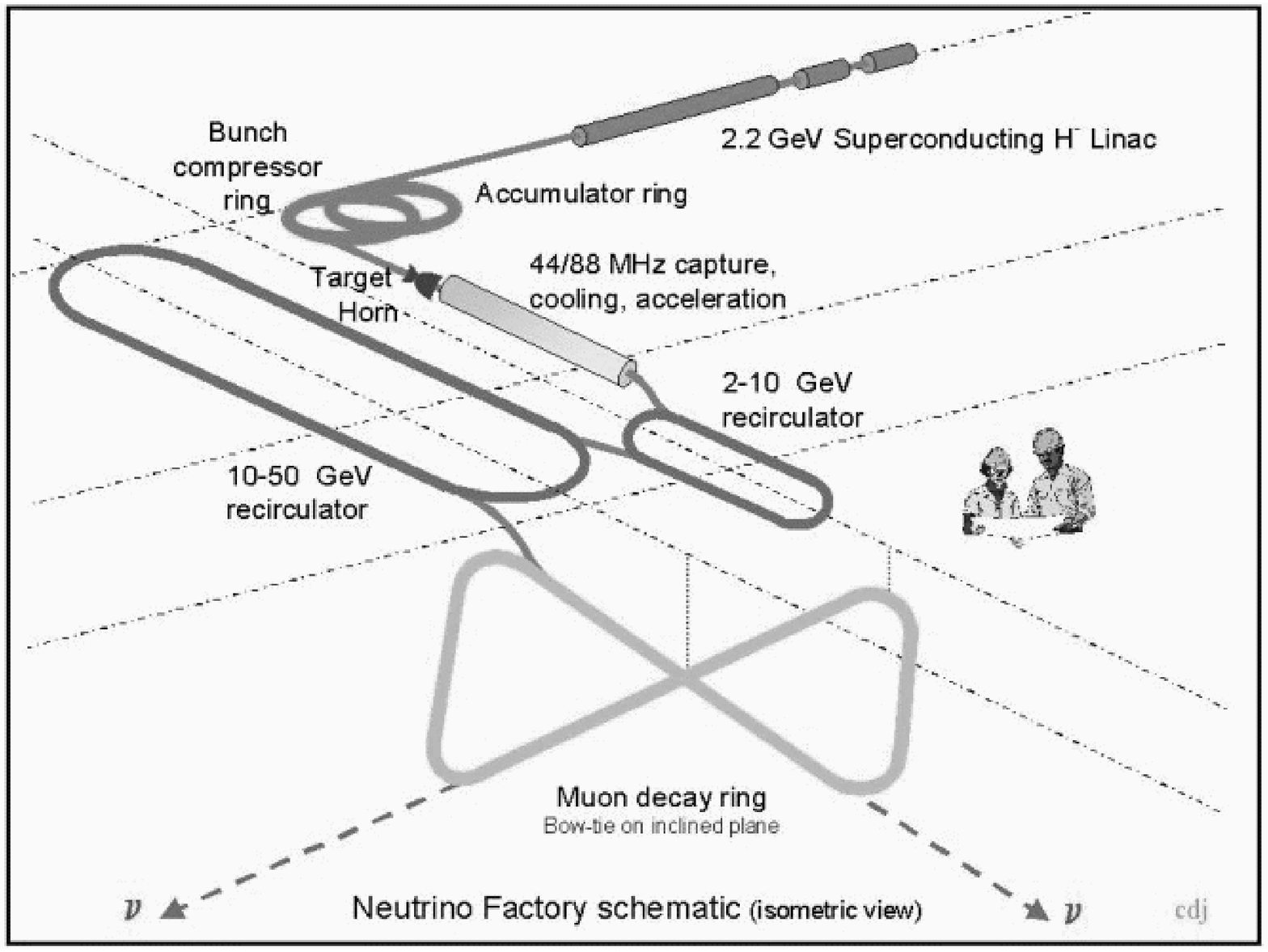,height=3in} 
\end{center}
\caption{{\it Conceptual layout of a $\nu$ factory at CERN, based on the
SPL~\protect\cite{SPL}.}}
\label{fig:nufactory}
\end{figure}

\section{Concept for a $\nu$ Factory}

A conceptual layout of a $\nu$ factory at CERN is shown in
Fig.~\ref{fig:nufactory}. The first requirement is an intense proton
source.  At CERN, the favoured option is a superconducting linac
(SPL)~\cite{SPL}, but higher-energy designs based on rapid-cycling
synchrotrons are being considered elsewhere~\cite{FNAL,BNL}. The SPL would
accelerate $H^-$ ions to 2.2 GeV at a repetition rate of 75 Hz, producing
$1.1 \times 10^{16}$ particles per second, corresponding to 4~MW mean beam
power.  Much of the acceleration in the SPL would be provided by reusing
LEP RF cavities, providing a considerable saving in development effort and
cost.  The SPL could be used in a shallow tunnel just outside the present
CERN boundary fence, with easy connections to the existing CERN
accelerator complex~\cite{SPL}. The target system would be non-trivial, in
view of the intense beam power.  Ideas for an external target include a
Mercury jet or a hot, rapidly-rotating, magnetically-levitating metal
band.

The SPL would provide better beams for the CERN PS accelerator, could
increase the current to the present isotope facility ISOLDE by a factor 5,
and perhaps feed a new radioactive beam facility.  Under study are the
possibilities of improved PS and SPS beams, e.g., for the CNGS $\nu$ beam,
and better beams (or at least a shorter filling time) for the LHC. Another
option could be a low-energy $\nu$ `superbeam' directed towards a
laboratory about 100 km away, e.g., near Modane at the location of the
Fr\'ejus tunnel~\cite{Frejus}. 

For a $\nu$ factory, the next step would be an accumulator/compressor ring
for manipulating the time structure of the SPL beam, which could be housed
in the old ISR tunnel. One would also need a horn system for focusing the
pions produced in the target and capturing the muons emanating from their
decays.  Next would be a complex system for muon cooling and phase
notation, designed to `tame' the muon beams.  This would be followed by a
set of recirculating linacs to accelerate the muons to the preferred
storage energy, between 20 and 50 GeV. They would then be transferred to a
tilted storage ring, with a triangular or bowtie shape, where they would
be allowed to decay, mainly in straight sections sending beams to
detectors at various baseline distances. 

Much more muon cooling would be needed for a muon collider, e.g., one
suitable as a factory to produce Higgs bosons (perhaps in the hopes of
studying CP violation in their decays), or a high-energy frontier $\mu^+
\mu^-$ collider~\cite{Hfactory}. We should not lose sight of the muon
collider as the prospective `faint blue dot' for the long-term development
of a muon storage ring complex. 

Other speakers at this meeting discuss in more detail the neutrino
oscillation physics opportunities opened up by a neutrino
factory~\cite{PH,others}. In the following sections, I discuss in more
detail some of the other physics opportunities offered by such a muon
complex. 

\section{Charged-Lepton-Flavour Violation and Muon Physics}

Even before accelerating the muons produced at the front end of a $\nu$
factory, there are many physics opportunities using slow and stopped
muons, one of the most interesting being the search for $\mu \rightarrow
e$ transitions.

If neutrino experiments are observing $\nu$ oscillations, should one also
expect observable transitions violating charged-lepton number, such as
$\mu \to e$, $\tau \to \mu$ or $\tau \to e$ transformations? If the only
modification to the Standard Model is to add heavy singlet (right-handed) 
neutrinos in some seesaw model (\ref{seesaw}), any such transitions would
be unobservably rare, since they would be suppressed by inverse powers of
the heavy Majorana mass $M_M$.  However, they could be observable in a
{\it supersymmetric} GUT model. Between the GUT scale and the heavy
neutrino mass scale $M_M$, the neutrino Yukawa couplings $\lambda_D$,
which in general are off-diagonal, renormalize the slepton and sneutrino
mass matrices:
\beq
\delta m^2_{\tau,\tilde{\nu}} \simeq
\frac{1}{8 \pi^2} \left( 3m^{2}_{0} + A_0^2 \right) \; \ell n 
\left ( \frac{m_{GUT}}{M_m} \right ) \;
\lambda^{+}_{D} \lambda_D ,
\label{8}
\eeq
where $m_0$ is the soft supersymmetry-breaking scalar mass~\footnote{We
assume this to be universal for the different lepton generations: if not,
the rates for charged-lepton-flavour violation would be further enhanced.}
and $A_0$
is the trilinear soft supersymmetry-breaking parameter.  Once off-diagonal
entries in $m^2_{\tau}$ and/or $m^2_{\tilde{\nu}}$ are induced, diagrams
similar to those responsible for $(g_\mu - 2)$ induce $\mu \to e$, $\tau
\to \mu$ and $\tau \to
e$ transitions, suppressed only by powers of $m^2_{\tau , \tilde{\nu}} \,
\lappeq 1$ TeV$^2$. 

\begin{figure}[htb]
\begin{center}
\epsfig{figure=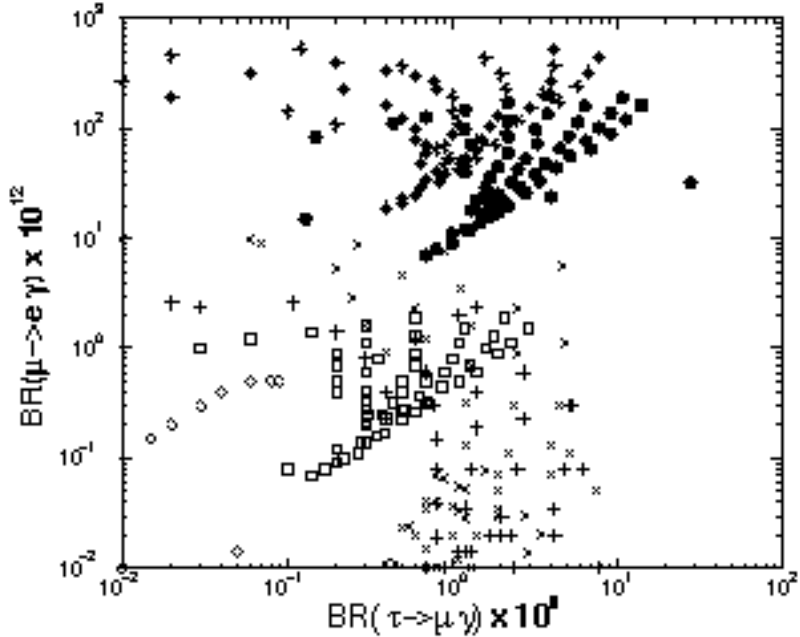}
\end{center}
\caption{{\it Scatter plot of predictions for $\mu \rightarrow e \gamma$
and $\tau \rightarrow \mu \gamma$ in a sampling of supersymmetric GUT
models with neutrino flavour textures motivated by the data on neutrino
oscillations~\protect\cite{EGLLN}.}}
\label{fig:EGLLNscatter}
\end{figure}

Fig.~\ref{fig:EGLLNscatter} shows rates for $\mu \to e \gamma$ and $\tau
\to \mu \gamma$ in representative models of fermion mass textures
motivated
by the neutrino oscillation data~\cite{EGLLN}.  We see that there may be
$\mu \to e$ decays at a rate within two orders of magnitude of the present
limit $B (\mu \to e \gamma )  < 1.2 \times 10^{-11}$, and it is also
possible that $\tau \to \mu \gamma$ might appear within two orders of
magnitude of the present limit $B (\tau \to \mu \gamma ) < 1.1 \times
10^{-6}$. 

Related to $\mu \to e \gamma$ are the processes $\mu \to 3e$, which are
expected to proceed mainly via $\gamma$ exchange with a rate
\beq
\frac{B (\mu \to 3e)}{b (\mu \to e \gamma )} \simeq
\frac{\alpha}{3 \pi} \;
\left [ \ell n (m^2_{\mu}/m^2_e) - \frac{11}{4} \right ]
\simeq 6 \times 10^{-3}
\label{9}
\eeq
and $\mu \to e$ conversion on a heavy nucleus such as Titanium.  The rate
for this process is typically suppressed relative to $\mu \to e \gamma$ by
a factor similar to (\ref{9}), but other diagrams may contribute and this
ratio is not universal. The SINDRUM II project at PSI hopes to reach a
sensitivity $B (\mu Ti \to e Ti) \sim 5 \times 10^{-13}$~\cite{SINDRUM},
the MECO project at BNL aims at $B (\mu Ti \to e Ti) \sim 5 \times
10^{-17}$~\cite{MECO}, and an experimentt at the JHF might be able to
reach
$10^{-18}$~\cite{PRISM}.  These experiments may well be sensitive to the
estimated rate of $\mu \to e$ conversion in a supersymmetric GUT. 

{
\begin{table}[bth]
\label{muon_experiments}  
\caption[]{\it
Experiments which could benefit from the intense
stopped muon sources at a $\nu$ factory~\protect\cite{GG}.}
\hspace*{-0.70in}
{\tiny \begin{tabular}[b]{|c|c||c|c|c||c|}
\hline
Type of   & Physics Issues & Possible   & previously established &present 
activities &projected for \\
Experiment&                & Experiments&accuracy&(proposed accuracy)&
NUFACT @ CERN \\
\hline \hline
''Classical'' & Lepton Number Violation;&$\mu^-N \to e^-N$
&$6.1 \times
10^{-13}$       & PSI, proposed BNL       ($5 \times 10^{-17}$) &  $ <
10^{-18}$     \\
Rare \&       & Searches for New Physics:&$\mu \to e \gamma$
&$1.2 \times 10^{-11}$
                & proposed PSI           ($2 \times 10^{-14}$) &  $ <
10^{-15}$      \\
Forbidden     & SUSY, L-R Symmetry,&$\mu \to eee$     &  $1.0
\times 10^{-12}$
                & completed 1985 PSI     & $ < 10^{-16}$ \\
Decays        & R-parity violation,.....&$\mu^+e^- \to
\mu^-e^+$&$8.1 \times 10^{-11}$
                & completed 1999 PSI &  $ < 10^{-13}$ \\
\hline
Muon          & $G_F$; Searches for New
Physics;&$\tau_{\mu}$                   &$18
\times 10^{-6}$         & PSI (2x), RAL         ($1 \times 10^{-6}$) &
$ < 10^{-7}$ \\
Decays         & Michel Parameters&$transv. Polariz.$
&$2\times 10^{-2}$& PSI, TRIUMF           ($5 \times 10^{-3}$) &   $ <
10^{-3}$ \\
\hline
&Standard Model Tests;&&&&\\
Muon           &   New Physics; CPT Tests
&$g_{\mu}-2$                    &$1.3 \times 10^{-6} $         & BNL 
($3.5\times10^{-7}$) &  $ <10^{-7}$        \\
Moments        &T- resp. CP-Violation &$edm_{\mu}$    &$3.4 \times
10^{-19} e\,cm$
            & proposed BNL           ($10^{-24} e\,cm$) &  $ < 5 \times
10^{-26} e\,cm$  \\
&in 2nd lepton generation&&&&\\
\hline
Muonium        & Fundamental Constants,
$\mu_{\mu}$,$m_{\mu}$,$\alpha$;&$M_{HFS}$ &$12 \times 10^{-9}$
              & completed 1999 LAMPF  &   $ 5 \times 10^{-9}$   \\
Spectroscopy   &  Weak Interactions; Muon Charge
&$M_{1s2s}$                     &$1 \times 10^{-9}$
                     &  completed 2000 RAL  &  $ < 10^{-11}$    \\
\hline
Muonic Atoms  & Nuclear Charge Radii;&$\mu^- atoms$ &$depends$
& PSI, possible CERN    &  new nuclear\\
&Weak Interactions&&&($<r_p>$to $10^{-3}$)& structure           \\
\hline
Condensed      & surfaces, catalysis & surface $\mu$SR             &$n/a$
& PSI, RAL             ($ n/a   $)& high rate                \\
Matter&bio sciences ... &&&&\\
\hline
\end{tabular}}
\end{table}}

\begin{figure}
\hspace*{-.70in}
\begin{minipage}{8in}
\epsfig{file=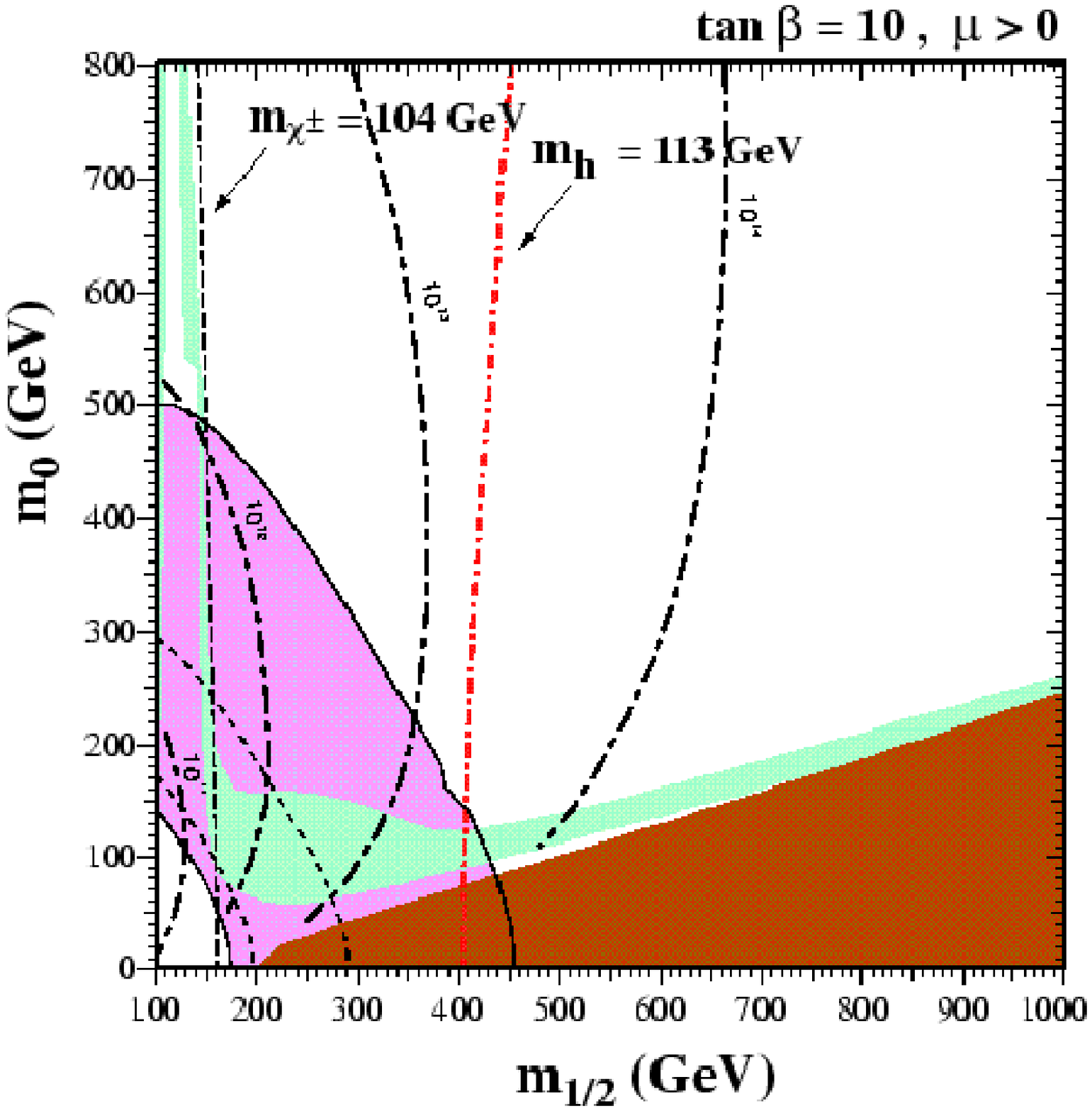,height=3.5in}
\epsfig{file=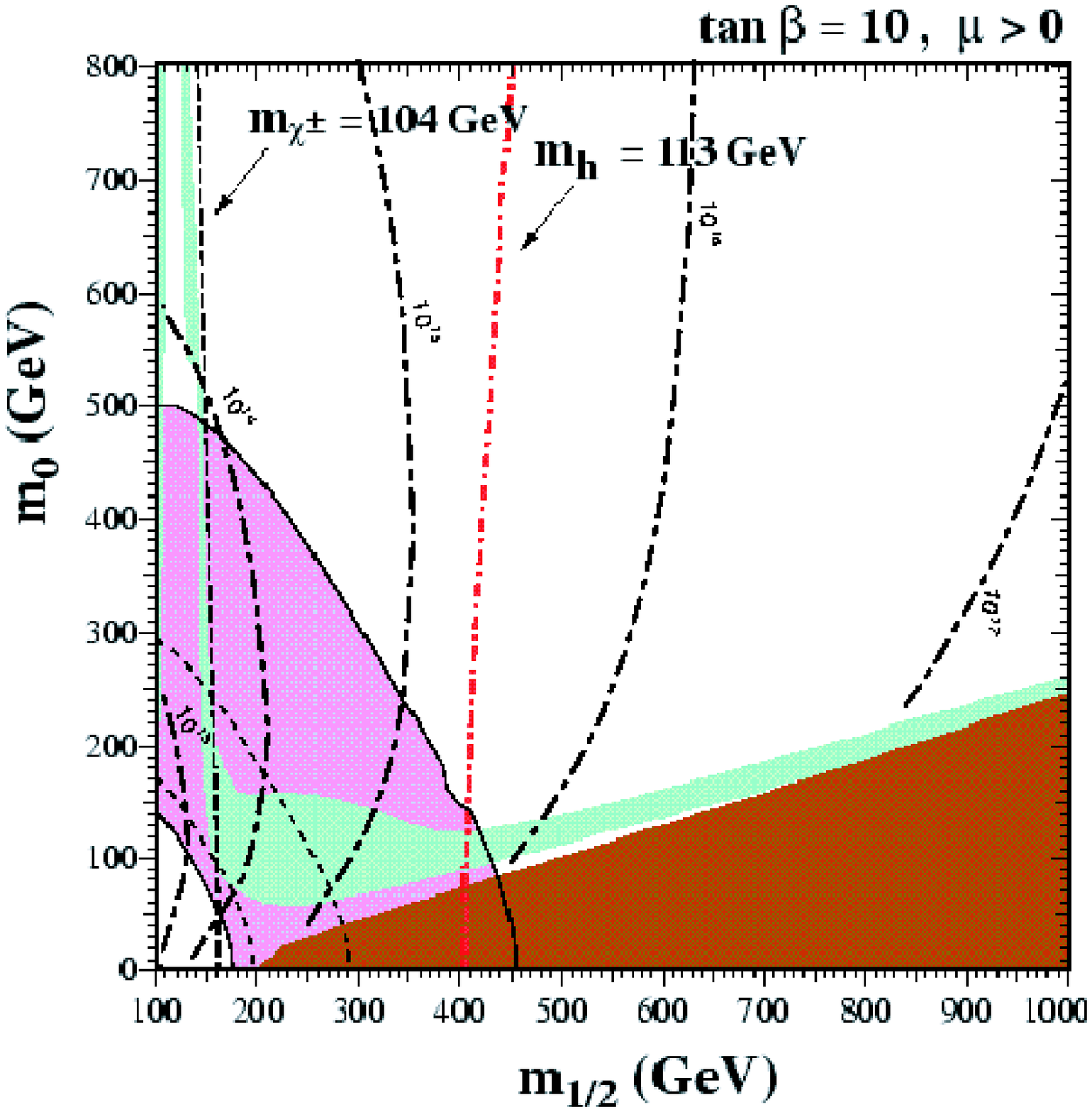,height=3.5in} \hfill
\end{minipage}
\caption{\label{fig:CEGL}
{\it Examples of the rates for (a) $\mu \rightarrow e \gamma$ and (b) $\mu
Ti \to e Ti$ in a supersymmetric GUT model with
$\tan \beta = 10$, $\mu > 0$ and one particular choice of flavour
texture for the neutrino mass matrices. The light (turquoise)
shaded areas are the cosmologically preferred regions, and in the dark
(brick
red) shaded
regions the LSP is the charged ${\tilde \tau}_1$, which is
excluded. The (pink) shaded regions are favoured by the measurement of
$g_\mu - 2$ at the 2- (1-)$\sigma$ level (solid and dotted
lines)~\protect\cite{CEGL}. }}
\end{figure}

Encouragement to search for $\mu \to e$ transitions has been provided by
the recent report of a possible experimental discrepancy with the Standard
Model prediction for the muon anomalous magnetic moment, $g_{\mu}
-2$~\cite{E821}.  Taken at face value, this suggests a non-trivial
flavour-diagonal $\mu^+ \mu^- \gamma$ vertex with an internal scale
$\lappeq 1$ TeV, corresponding to the appearance of new physics at this
scale. Neutrino oscillations suggest that the flavour-diagonal $\mu^+
\mu^- \gamma$ vertex should be accompanied by the corresponding
flavour-off-diagonal $\mu^{\pm} e^{\pm} \gamma$ vertex. A concrete example
of this expectation is provided by supersymmetry, as discussed
above~\cite{CEGL}. Ideally, one uses $(g_{\mu} - 2)$ to predict the
sparticle mass scale and $\nu$ oscillation data to quantify the flavour
mixing~\cite{CEGL}.  In practice, there are ambiguities in both these
steps.  However, as seen in Fig.~\ref{fig:CEGL}, there is reason to think
that $\mu \to e$ transitions might appear within two (or three)  orders of
magnitude of the present limits if the $(g_{\mu} - 2)$ discrepancy
eventually stabilizes within the present one- (two-)  $\sigma$ range. 

There are many other physics opportunities with an intense low-energy muon
source~\cite{GG}, notably including a follow-up experiment on $(g_{\mu}
-2)$ itself! Another interesting search could be that for a muon electric
dipole moment $d^e_\mu$, as a probe of CP violation. In many models, this
is
expected to be enhanced relative to that of the electron by a factor
$m_{\mu}/m_e \sim 200$, so a sensitivity better than $10^{-25}$ e.cm would
have greater physics reach than the present limit on the electric dipole
moment of the electron.  Muonium-antimuonium conversion is also a
possibility, and refined experiments measuring the muon lifetime and
Michel decay parameters would be interesting. Looking beyond experiments
of interest to particle physicists, one should also mention experiments on
muonic atoms, and the use of muons as probes in condensed-matter physics
and the biological sciences. 

\section{Deep-Inelastic Scattering}

\begin{figure}[htb]
\begin{center}
\epsfig{figure=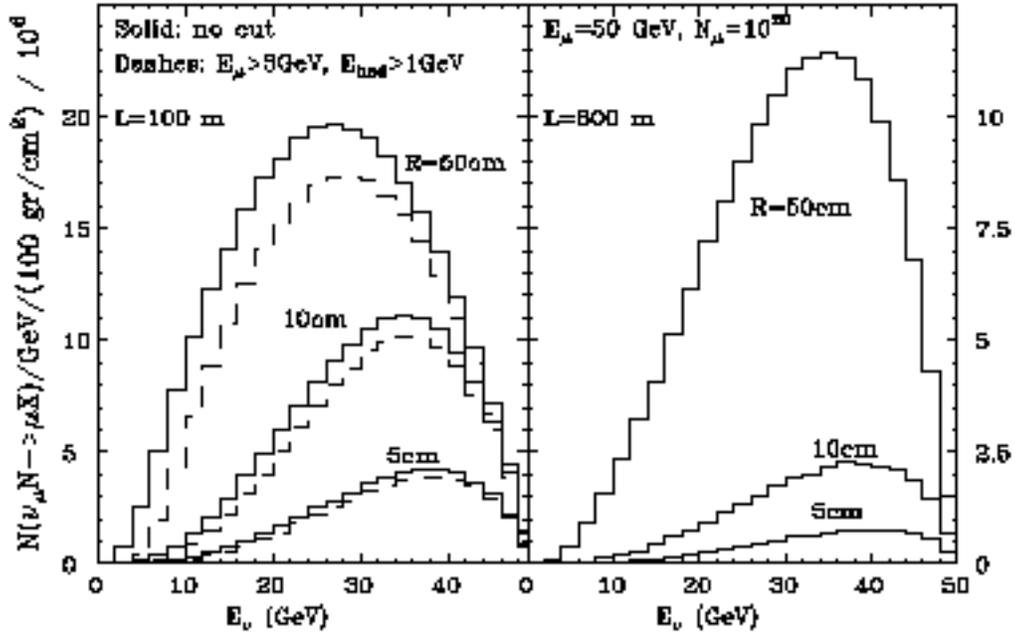}
\end{center}
\caption{{\it Examples of the $\nu$ deep-inelastic sacttering rates
possible in short-baseline experiments at a $\nu$
factory~\protect\cite{MM}.}}
\label{fig:NUDIS}
\end{figure}

Further new particle physics opportunities arise once muons are
accelerated and stored in a `ring'.  Many of the neutrinos produced in the
decays of stored muons may be directed towards a short-baseline target and
used in conventional deep-inelastic scattering experiments~\cite{MM}.
Experiments with over $10^8$ events in the $(x,y)$ plane are feasible, as
seen in Fig.~\ref{fig:NUDIS}, enabling `difficult' combinations of parton
distribution functions, such as $s(x) - \bar{s}(x)$, to be measured
directly for the first time.  The beams are so thin and intense that
relatively small detectors will have large rates, opening options such as
a polarized target or a silicon vertex detector.  There are five
measurable structure functions in polarized $\nu p$ scattering, and
$g_1^{W^+ + W^-}$ measures directly the singlet combination $\Delta n +
\Delta d + \Delta s (+ \Delta c)$, whilst $(g_5)_{p+n}^{\nu - \bar{\nu}} =
\Delta s (- \Delta c)$.  Using a silicon vertex detector, one will have a
much better handle on heavy quark production, enabling measurements of the
CKM matrix elements $V_{cs}$ and $V_{cd}$ to be improved. The high
statistics will also reduce greatly the statistical errors in the
determination of $\alpha_s (m_Z)$, to $\pm 0.0003$, and in $\sin^2
\theta_W$, to $\pm 0.0002$~\cite{MM}. 

Another possibility that should be investigated is deep-inelastic $\mu N$
scattering using an internal target in the muon storage ring.  The
kinematic range would be similar to that of the proposed ELFE
accelerator~\cite{ELFE}.  Are there interesting aspects of ELFE physics
that could be attacked using muons? 

\section{High-Intensity Kaon Physics}

Large numbers of kaons could be produced with a high-energy ($E \gappeq
15$ GeV) proton driver for a neutrino factory~\footnote{Such a
higher-energy driver has been favoured in some studies~\cite{FNAL,BNL}.},
or by
constructing a post-accelerator for some fraction of the protons from an
intense low-energy driver.  Several interesting physics opportunities
would be presented by intense kaon beams~\cite{{K0pi0}}.

$K^0_L \to \pi^0 \bar{\nu} \nu$~\cite{K0pi0}: This decay
measures the CP-violating combination $Im V^0_{ts} V_{td}$ of CKM matrix
elements, and complements the measurements of $\sin 2 \beta$ at B
factories.  In the Standard Model, one expects $B(K^0_L \to \pi^0
\bar{\nu} \nu) = (2.8 \pm 1.1) \times 10^{-11}$, and one could hope for a
$10\%$ measurement using a K factory option for a neutrino factory. 

$K^+ \to \pi^+ \bar{\nu} \nu$~\cite{Kpi}: This decay measures
$|V_{td}|$, and in combination with the neutral mode would determine $\sin
2 \beta$ with an error of $\pm 0.07$ in the Standard Model.  This would
therefore be the sensitivity to extensions of the Standard Model. 

$K^0_L \to \pi^0 e^+ e^-$~\cite{Kpiee}: This interesting decay
has contributions from direct CP violation $\propto Im V^*_{ts}V_{td}$
(which may be measurable at the $10\%$ level), indirect CP violation
related to $K_s \to \pi^0 e^+ e^-$, and the CP-conserving mechansim $K_L
\to \pi^0 \gamma^* \gamma^* \to \pi^0 e^+ e^-$.

\underline{$K^0_L \to \mu^{\pm} e^{\mp}$}~\cite{Kmue}: This process
violates both lepton and quark flavour, and may arise from box diagrams in
the presence of slepton and/or sneutrino mixing, which is expected in
supersymmetric GUT models of neutrino masses.  It is possible that
$B(K^0_L \to \mu^{\pm} e^{\mp}) \sim 10^{-18}$, but this would not be an
easy mode to observe at this level.

\section{Concluding Remarks}

Long-baseline neutrino oscillation experiments are the `core business' of
a neutrino factory.  They bear directly upon the fundamental issues of
flavour and unification, and offer many fascinating possibilities for
interesting experiments. 

In addition, neutrino factories open the way to many other exciting
projects, such as muon colliders used either as Higgs factories or at the
high-energy frontier. 

Moreover, there are many interesting physics opportunities which could be
exploited earlier, for example in low-energy muon physics using the front
end of a neutrino factory: $\mu \to e \gamma$, $\mu \to 3 e$, $\mu \to e$
conversion on nuclei, $g_{\mu} -2$, $d^e_{\mu}$, and many more.  Once the
produced muons are accelerated, new opportunities such as deep-inelastic
$\nu N$ and $\mu N$ scattering are made available. Intense higher-energy
proton beams would also offer interesting possibilities in kaon physics.

A neutrino factory will be a complex and expensive project. Whilst
neutrino oscillation physics is its primary motivation, neutrino
physicists should recognize that they are a minority of the
particle-physics community. We are likely to need the support and active
involvement of other communities if a neutrino factory project is to be
realized.  For this reason, we must work together with these other
communities of physicists for the successful realization of a neutrino
factory.

\section{Acknowledgements}
It is a pleasure to thank Alain Blondel, Friedrich Dydak, Helmut Haseroth
and other members of the Muon Steering Group for many valuable
discussions. I also thank Gerhard Buchalla, Gian Giudice, Pilar Hernandez
and Michelangelo Mangano for their input on several aspects of
the physics discussed here. Finally, I think Milla Baldo-Ceolin for her
kind invitation to an excellent meeting in very pleasant surroundings.

\end{document}